# Power Control for 6G Industrial Wireless Subnetworks: A Graph Neural Network Approach

Daniel Abode, Ramoni Adeogun, Gilberto Berardinelli
*Department of Electronic Systems, Aalborg University*
Aalborg, Denmark
Email:{danieloa, ra, gb}@es.aau.dk

*Abstract*—6th Generation (6G) industrial wireless subnetworks are expected to replace wired connectivity for control operation in robots and production modules. Interference management techniques such as centralized power control can improve spectral efficiency in dense deployments of such subnetworks. However, existing solutions for centralized power control may require full channel state information (CSI) of all the desired and interfering links, which may be cumbersome and time-consuming to obtain in dense deployments. This paper presents a novel solution for centralized power control for industrial subnetworks based on Graph Neural Networks (GNNs). The proposed method only requires the subnetwork positioning information, usually known at the central controller, and the knowledge of the desired link channel gain during the execution phase. Simulation results show that our solution achieves similar spectral efficiency as the benchmark schemes requiring full CSI in runtime operations. Also, robustness to changes in the deployment density and environment characteristics with respect to the training phase is verified.

## I. INTRODUCTION

The flexible and reconfigurable manufacturing vision of Industry 4.0. [1] relies on the availability of highly reliable wireless connectivity at the field level to replace rigid wired connections [2]. This motivated a novel radio concept termed in-X subnetworks, short-range cells consisting of controllers acting as the access point to some sensors and actuators for field-level control, described as Industrial Wireless Subnetworks (IWS) in this paper. Such subnetworks are envisioned as a relevant component of upcoming $6^{\text{th}}$ Generation (6G) systems [3], [4]. They can be installed in robots or production modules to support local connectivity for critical industrial use cases [4]–[6], such as the closed-loop control systems and safety systems [7]. Besides, an industrial scenario may require highly dense deployment of subnetworks to cater for the large number of sensors, actuators and controllers for a diverse set of applications and use cases. This would heighten interference, thereby limiting the network's spectral efficiency (SE). Interference reduction techniques such as power control are therefore crucial for a spectrally efficient IWS deployment. Several works (e.g., [8]–[14]) have proposed efficient power control methods that however require input measurements of the inter-cell interference levels reported by each cell. Acquiring such information introduces significant delays and large signalling overhead. In addition, the computational complexity of the power control algorithm can be significantly high, especially in the case of a high number of cells. The practicality of efficient power control for a system such as IWS depends on resolving these limitations.

The quest to achieve efficient power control with low computational complexity has influenced the adoption of neural network (NN) methods to replace traditional complex iterative algorithms such as Geometric programming (GP) [8] and Weighted Minimum Mean Square Error (WMMSE) [9]. Initial works [10], [11] used feed-forward NN, but the NN architecture can not adapt to changes in the wireless network, such as changes in the number of links. Recent works [12]–[14] employed a novel NN technique, message passing graph neural networks (MPGNN) leveraging the wireless network topology in the form of a graph model as an input. They showed that the MPGNN approach offer advantages such as invariance to network topology, lower complexity, better robustness and performance compared to feed-forward NN methods. However, the aforementioned methods still depend on the full channel state information (CSI) as input data, but in practice, obtaining the full CSI introduces considerable delays. Meanwhile, for IWS, the capability of prompt decision is of fundamental importance as some in-robots subnetworks may be mobile causing inter-subnetworks interference levels to vary rapidly over time [6]. In this respect, delays in acquiring full CSI can be a considerable limitation.

In this paper, we propose a centralized power control solution for IWS based on MPGNN that only requires limited CSI knowledge in runtime operations. IWS are expected to operate within the coverage of a central network, hence a central resource manager (CRM) can be employed for power control. While the training phase relies on full CSI acquisition, in the execution phase, only information on the geometrical distance between subnetworks and the desired link CSI is needed at the CRM. Since information on the mutual distance among subnetworks can be known in advance (or planned by the CRM itself), signalling is therefore reduced to the desired CSI, avoiding the necessity of sensing and reporting interference. The CRM uses such information for updating a graph representation of the IWS deployment consisting of the desired link CSI as the node features and the interfering links'

This project has received funding from the European Union's Horizon 2020 research and innovation programme under the Marie Skłodowska-Curie grant agreement No 956670 geometrical distances as the edge features. Such graph representation serves as the input for the MPGNN power control algorithm. We show that the proposed method achieves similar performance as the methods using complete CSI. Moreover, it is equally robust to changes in deployment parameters such as shadowing standard deviation or subnetworks density.

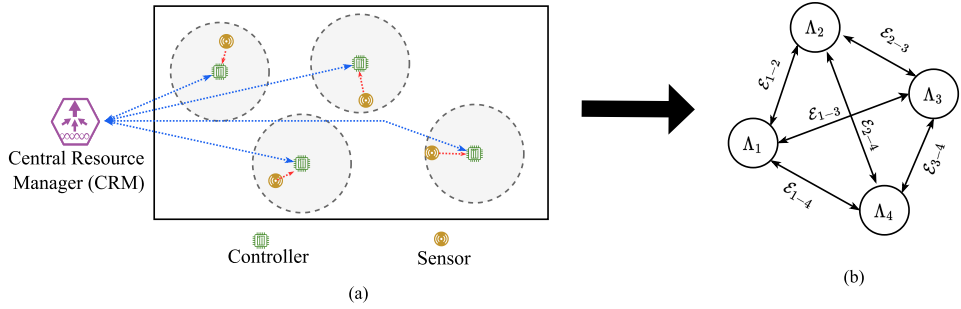

Fig. 1. Simple representation of IWS. (a) Each cell consists of a controller and uplink communication between the sensor and controller. Signalling links exist between the subnetworks' controllers and a central resource manager (b) Graph model of the subnetworks. Note that the graph is a directed graph, the edge notation $\mathcal{E}_{n-m}$ implies both $\mathcal{E}_{n,m}, \mathcal{E}_{m,n}$.

The remainder of this paper is structured as follows. We introduce our proposed graph model of industrial wireless subnetworks in Section II, while Section III describes the proposed graph-driven power control method. In Section IV, we evaluate the performance of the proposed method. Finally, we summarise our observations and present directions for future work in Section V. The code for the numerical evaluation is publicly available on GitHub[1].

## II. GRAPH MODEL OF INDUSTRIAL WIRELESS SUBNETWORKS

IWS are short-range cells installed in robots, production modules, automated guided vehicles (AGVs), conveyors and other industrial machines, consisting of an access point with controller capabilities and associated sensors and actuators. Fig. 1 shows a simplified representation of a 2D layout of an IWS deployment in a factory environment. The representation shows a single uplink between a sensor and a controller in each subnetwork, and a signalling link from each subnetwork's controller to a central resource manager (CRM). It is assumed that all the devices within a subnetwork are allocated orthogonal resources making inter-cell interference the main limitation to the subnetwork's spectral efficiency (SE) [6]. We focus on the uplink transmission in a network of $N$ subnetworks. For simplicity, for the rest of the paper, we assume that each subnetwork serves a single device whose transmissions occupy the available bandwidth. In subnetwork $n$, the device is located at coordinate $(x_n, y_n)$ and the controller at the centre of the subnetwork $(x_n^0, y_n^0)$. The channel gain of the desired link in subnetwork $n$, which comprises the large-scale path loss, shadowing and small-scale fading is denoted as $h_{n,n} \in \mathbb{R}$. In the same way, the channel gain of the interfering link between the device in subnetwork $n$ and the access point in subnetwork $m$ is denoted as $h_{n,m}$, and the corresponding distance is denoted as $d_{n,m} \in \mathbb{R}$, $n, m \in \{1, 2, .., N\}$. Accordingly, we formulate an attributed graph model $G(\Lambda, \mathcal{E}, \Gamma)$ of the deployment as shown in Figure 1b, where $\Lambda$ is a finite set of nodes corresponding to the subnetworks and $\mathcal{E} \in \mathbb{B}^{N \times N}$ is an adjacency matrix of booleans indicating the presence or absence of edges. $\Gamma \in \mathbb{R}^{\mathbb{N} \times \mathbb{N}}$ represents a matrix of the graph attributes, so that;

[1]https://github.com/danieloaAAU/Power_Control_GNN.git

$$\Gamma_{n,m}^{hD} = \begin{cases} h_{n,n}, & \text{if } n = m, \\ d_{n,m}, & \text{if } n \neq m. \end{cases} \quad (1)$$

To build the graph, the CRM only requires information of the desired channel, $h_{n,n}$ from the subnetworks since all the subnetworks' positioning coordinates and as such their mutual distances can be already available at the CRM. In the case of mobile subnetworks, e.g. those installed in AGVs, the CRM can be aware of the robot routes and therefore the expected mutual distances over time through its knowledge of the factory's operations plan. The CRM would use the graph representation as an input to an MPGNN algorithm to optimize power allocation. The next section presents the proposed architecture for power control.

## III. UNSUPERVISED CENTRALIZED POWER CONTROL ARCHITECTURE

In this section, we describe the architecture of the proposed power control method based on the MPGNN algorithm on a network graph attributed by $h_{n,n}$ and $d_{n,m}$ as nodes and edges features respectively. We refer to our method as power control graph neural network –hD (PCGNN-hD), where the acronym "hD" indicates that the desired link channel gain, "h" and the interfering link distances, "D" are used as the graph features. Fig. 2 shows the training and inference process, which is further described below.

For our power control problem, the objective is to decide the array of transmit powers $\mathbf{P} \ni p_n$ for all subnetworks that maximize the network's sum SE, $\Upsilon(\mathbf{P}, \mathbf{H})$ given the channel gain matrix $\mathbf{H} \in \mathbb{R}^{N \times \mathbb{N}}$, i.e.

$$\begin{aligned} \underset{\mathbf{P}}{\text{maximize}} \quad & \Upsilon(\mathbf{P}, \mathbf{H}), \\ \text{subject to} \quad & \\ & 0 \leq p_n \leq P_{max} \ \forall p_n \in \mathbf{P}, \end{aligned} \quad (2)$$

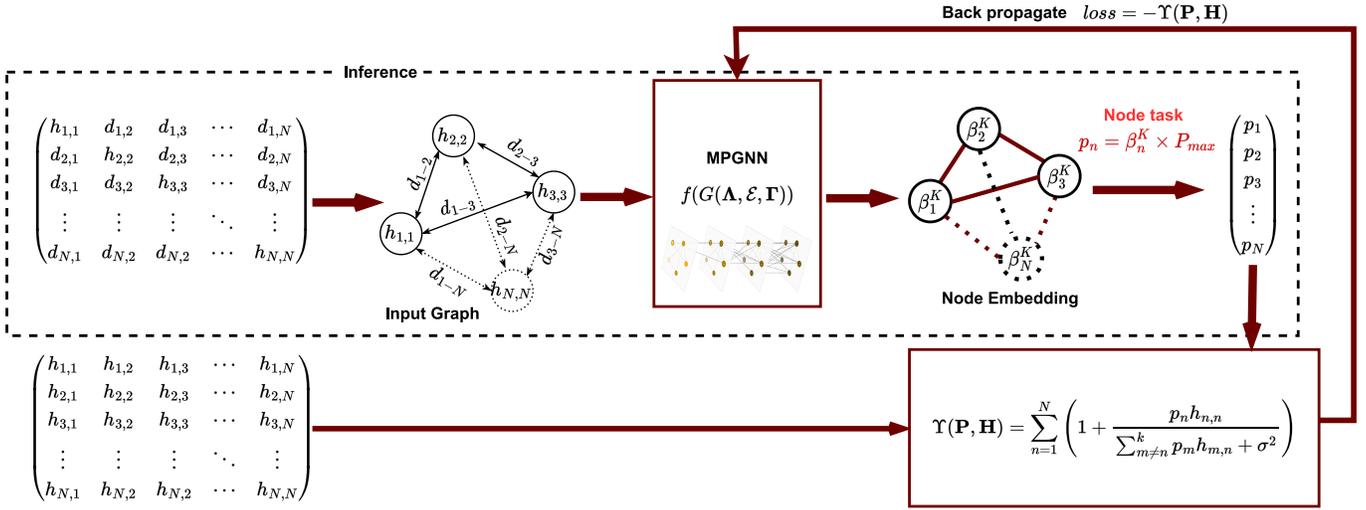

Fig. 2. The architecture of PCGNN-hD, a centralized power control based on the Message Passing Graph Neural networks (MPGNN) algorithm with an input graph attributed by desired links CSI and interfering links geometric distances. The ticked box highlights the inference procedure while all parts of the architecture are involved in the training procedure.

where

$$\Upsilon(\mathbf{P}, \mathbf{H}) = \sum_{n=1}^{N} \log_2 \left(1 + \frac{p_n h_{n,n}}{\sum_{\substack{m=1 \\ m \neq n}}^{N} p_m h_{m,n} + \sigma^2}\right). \quad (3)$$

In (3), $\sigma^2 = JTB \cdot 10^{NF/10}$ is the thermal noise power, where $J$ is the Boltzmann constant, $T$ is the temperature (kelvin), $B$ is the bandwidth (Hz), and $NF$ is the Noise figure (dB).

The MPGNN algorithm is a method of deep learning on graphs that involves the creation of messages at each node using the node as well as the edge attributes, and the exchange of the messages among neighbouring nodes [15]. The message passing framework consists of two operations; the aggregation function and the combination function, which form a layer of message passing computation. At the $k$-th layer, the aggregation function at each node is an arbitrary permutation invariant function that accumulates the neighbouring messages, $\beta_{\mathcal{N}(n)}^k$, where $\mathcal{N}(n)$ refers to the neighbours of the subnetwork $n$. The combination function combines the aggregation output, $\beta_{\mathcal{N}(n)}^k$, with the previous embedding of the node, $\beta_n^{k-1}$ to produce a new embedding for the node $n$, $\beta_n^k$ at layer $k$. Our adaptation of the MPGNN algorithm for power control, PCGNN-hD is shown in Algorithm 1.

Some mathematical and neural network functions have been proposed for both aggregation and combination. We chose the mean aggregation function (denoted as Mean()), and we parameterized the message computation and combination with artificial neural networks (ANN) to generate trainable message computation weights, $\mathbf{W}$ and message combination weights, $\mathbf{Z}$. According to earlier works on MPGNN [12]–[15] using the ANN for message computation and the combination helps to improve the expressive and generalization ability of the MPGNN. Note that in Algorithm 1, the symbol $\parallel$ denotes vector concatenation.

---

**Algorithm 1:** PCGNN-hD

**Input:** Graph $G(\mathbf{\Lambda}, \mathcal{E}, \mathbf{\Gamma})$; Number of subnetworks, $N$; Number of MPGNN layers, $K$; Message computation weights, $\mathbf{W}$; Message combination weights $\mathbf{Z}$; Rectified Linear Unit, $\sigma(\cdot)$; Sigmoid function $\zeta(\cdot)$; Aggregation function, Mean($\cdot$);
**Output:** $p_n \ \forall \ n \in 1, ..., N$;
**Initialize:** $\beta_n^0 \leftarrow h_n \leftarrow \mathbf{\Gamma}_{n,n} \ \forall \ n \in 1, ..., N$;
**Initialize:** $d_{n,m} \leftarrow \mathbf{\Gamma}_{n,m} \ \forall \ n \neq m; n, m \in 1, ..., N$;
**for** $k := 1$ *to* $K$ **do**
 **for** $n := 1$ *to* $N$ **do**
  $\beta_{\mathcal{N}(n)}^k \leftarrow \text{Mean}(\sigma(\mathbf{W}[\beta_m^{k-1} \parallel d_{m,n}]))$ ;
  $\beta_n^k \leftarrow \zeta(\mathbf{Z}[\beta_n^{k-1} \parallel \beta_{\mathcal{N}(n)}^k])$ ;
 **end**
 **if** $k \neq K$ **then**
  $\beta_n^k \leftarrow \beta_n^k \parallel \beta_n^{k-1}$;
 **end**
**end**
$p_n \leftarrow \beta_n^K \times P_{max}$

---

As discussed in Section II, the CRM updates the graphical representation of the subnetworks deployment as an input graph for the PCGNN-hD algorithm. The algorithm generates a node embedding for each subnetwork following the message passing framework. Finally, the transmit power is computed via a node task, $p_n = \beta_n^K \times P_{max}$ for all nodes. As shown in the construction of the input graph in Fig. 2, the only radio information needed for inference is the channel gain of the desired links which can be easily obtained. This is one of the advantages of this approach. However, during the training phase, the full CSI is needed to compute the network performance $\Upsilon(\mathbf{P}, \mathbf{H})$. We recommend that the training procedure is done offline due to the complexity of obtaining the full CSI. For example, a well-calibrated simulator or digital twin of the operational environment, eventually aided by preliminary measurement campaigns can generate the full CSI needed for the offline training process.

The loss function is defined as the negative of the network sum SE ($-\Upsilon(\mathbf{P}, \mathbf{H})$). During the training, the computed loss is backpropagated through the layers of the PCGNN-hD to update $\mathbf{W}$ and $\mathbf{Z}$ at every time step using gradient descent until convergence after a fixed number of iterations. The evaluation of the performance and robustness of the described method is presented in the next section.

## IV. NUMERICAL EVALUATION

### A. System model

We consider an IWS deployment of $N$ subnetworks in an $L \times L$ (m$^2$) factory area. Subnetwork $n$ has one controller in the centre of a circular cell of radius $R = 2$ m, and one sensor randomly positioned at distance $d_{n,n}$ with minimum proximity of 0.5 m to the controller. The channel gain in the link between the sensor at subnetwork $m$ and the access point in subnetwork $n$ is given by

$$h_{m,n} = \frac{c^2}{(4\pi f)^2 d_{m,n}^r} \cdot \kappa_{m,n} \cdot |\zeta_{m,n}|^2, \quad (4)$$

where $r$ denotes the path loss exponent, shadowing is denoted by $\kappa$ and small-scale fading as $\zeta$. Note that $c$ is the speed of light in (m/s), $f$ is the carrier frequency in (Hz), and $\kappa$ is randomly sampled from a lognormal distribution with standard deviation, $\lambda$; $\kappa \sim Lognormal(0, \lambda^2)$ for all links. Likewise, $\zeta$ is randomly sampled from a complex-valued normal distribution; $\zeta \sim \mathcal{CN}(0, 1)$. For the desired links, $n = m$ and for the interfering links, $n \neq m$, $n, m \in \{1, 2, .., N\}$. The achievable SE (bits/s/Hz) at subnetwork $n$ is approximated using the Shannon capacity equation as shown below:

$$C_n \approx \log_2 \left( 1 + \frac{p_n h_{n,n}}{\sum\limits_{\substack{m=1 \\ m \neq n}}^{N} p_m h_{m,n} + \sigma^2} \right). \quad (5)$$

For the dataset, we collected 10,000 snapshots for training and 50,000 snapshots for testing using different random seeds. Each snapshot is a different realization of the system model of the subnetworks' deployment. Hence the SE, shadowing values, distance, and channel gains vary randomly across all the links in a snapshot, and across the snapshots. The value of the main parameters used in our simulations is collected in Table I.

### B. PCGNN-hD Training specification

We consider a PCGNN-hD of $K=3$ layers; the message computation function is a 4-layer ANN with a fully connected structure of [2, 32, 32, 32] neurons with rectified linear units for activation. Likewise, the message combination function is a 4-layer ANN with a fully connected structure of [34, 32, 16, 1] neurons using the sigmoid activation function, which ensures that $\beta_n^k \in [0, 1]$. The size of the PCGNN-HD and the ANN is fine-tuned to offer the best performance from simulation trials. The weights of both ANN were initialized from a uniform distribution according to the method described in [17] and were updated using the adaptive moment estimation gradient descent algorithm (ADAM). The training was done for 1500 epochs which ensured the convergence of the PCGNN-hD as observed from the simulation trials.

TABLE I
SYSTEM ASSUMPTIONS

| Parameter | Value |
|---|---|
| Subnetwork radius, $R$ | 2 m |
| Number of devices per subnetwork | 1 |
| Minimum distance between controllers | 2 m |
| Sensor to controller minimum distance | 0.5 m |
| Factory area, $L \times L$ | 20 m x 20 m |
| Number of subnetworks per snapshot, $N$ | 20 |
| Deployment density (Subnetworks/$km^2$) | 50000 |
| Shadowing standard deviation, $\lambda$ | 7 dB |
| Path loss exponent, $r$ | 2.7 |
| Maximum transmit power, $P_{max}$ | 0 dBm |
| Bandwidth, $B$ | 20 MHz |
| Center frequency, $f$ | 6 GHz |
| Noise figure, NF | 10 dB |

### C. Benchmarks

We evaluate the achieved average SE and robustness of our proposed method in comparison with the following schemes.

- **Maximum power (Max Power):** The transmit power for all the links in the deployment is set to the maximum power, $P_{\max} = 0$ dBm, i.e. power control is not used. This was used as a benchmark in [10].
- **Weighted Minimum Mean Square Error (WMMSE):** This is a popular benchmark scheme for power control, used in [11], [14] and also as a ground-truth to supervise NN methods for power control in [10], [13]. It is an iterative method that requires the full CSI information input and has a high computational complexity, rendering it impractical for real-time applications [14]. We implemented WMMSE as in [10] to solve the objective function (2), with an option to terminate execution when the difference between the achieved sum SE at iteration $t$ and $t - 1$ is less than $10^{-5}$, which we found to take on average 10 iterations.
- **PCGNN-dD :** It employs the same PCGNN algorithm as in PCGNN-hD, but the attribute of the input graph $G(\mathbf{\Lambda}, \mathcal{E}, \mathbf{\Gamma})$ is based only on the distance of the desired links and interference links, i.e.,

$$\Gamma_{n,m}^{dD} = d_{n,m}, n, m = 1, 2, .., N. \quad (6)$$

The usage of geometric information only as input feature for power control was considered in [18].

- **PCGNN-hH :** It also uses the same PCGNN algorithm as *PCGNN-hD*, but the input graph $G(\Lambda, \mathcal{E}, \Gamma)$ is attributed by the full channel gain matrix as in

$$\Gamma_{n,m}^{hH} = h_{n,m}, n, m = 1, 2, .., N. \quad (7)$$

Using the full channel gain matrix as the input features is a common approach for neural network-based power control methods [10]–[14].

### D. Average SE performance

We evaluated the resulting average network SE over 50,000 test snapshots. Fig. 3 shows the Cumulative Distribution Function (CDF) of the average SE. We can observe that the *PCGNN-hD* improved the achievable average SE by 33% with respect to *Max Power*, with only a 3% penalty with respect to *PCGNN-hH* and a 2% gain with respect to *WMMSE*. *PCGNN-hH* and *WMMSE* both require the full channel gain matrix, but *PCGNN-hH* performs 5% better than *WMMSE*. The advantage of our method is better appreciated by looking at the CDF of the average SE achieved by *PCGNN-dD*. On average, *PCGNN-dD* achieved a gain of 15% compared to *Max Power*. In essence, even though the CRM has the knowledge of the location of the devices in the subnetwork, this information alone is not enough to efficiently maximize the SE. But, by combining the interfering distance information and the desired link channel gain as in *PCGNN-hD*, the gain in spectral efficiency was only 3% lower than *PCGNN-hH*.

Furthermore, in Fig. 4, we show the CDF of the average power allocated to the subnetwork deployment snapshots. We can observe that *PCGNN-hD* used marginally less power than *PCGNN-hH* as seen in the tail of the distribution. And on average 50% less power than *WMMSE*. Generally, the *PCGNN* methods tend to reduce the average power while improving the SE, hence translating to improved energy efficiency.

### E. Robustness Analysis

*PCGNN* methods are expected to be trained with data collected from an environment which closely resembles the operational environment. However, some differences are likely to occur during operation such as changes in the shadowing standard deviation, or changes in the deployment density. It is desirable that *PCGNN-hD* remain robust to these changes to ensure continuity of efficient operation without the need for a time-expensive retraining procedure. In this subsection, we sought to observe the robustness of the *PCGNN* methods to such changes.

1) **Shadowing standard deviation** $\lambda$**:** We consider three shadowing standard deviation $\lambda$ = 4 dB, 7 dB, and 10 dB as shown on the x-axis of Fig. 5(a) which displays the SE gain of the discussed PCGNN solutions with respect to Max power. We compared two cases; a) The network which is trained in

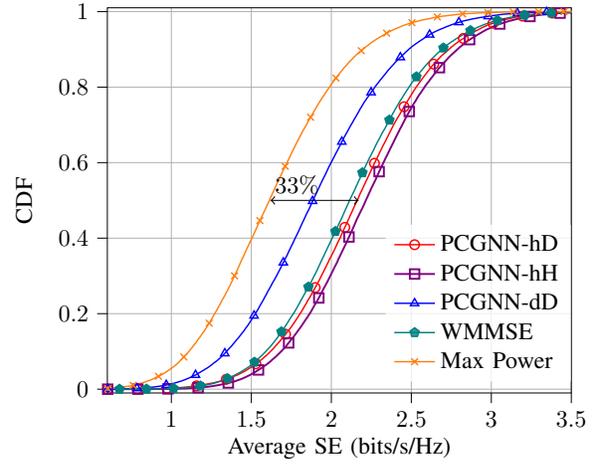

Fig. 3. Cumulative distribution function of the average SE per deployment over 50000 test snapshots

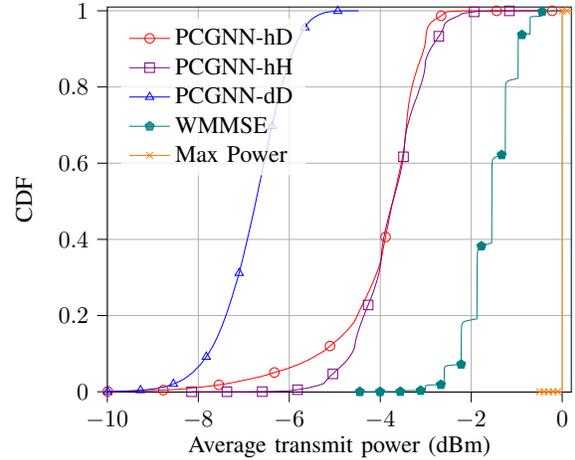

Fig. 4. Cumulative distribution function of the average power allocated per deployment over 50000 test snapshots

an environment with $\lambda$ = 7 dB is tested with three shadowing standard deviations $\lambda$ = 4 dB, 7 dB, and 10 dB. This is shown with the blue continuous lines (legend T1). b) We trained and tested the network in environments with the same shadowing standard deviation. This is shown by the red dashed line (legend T2). The positive slope of the lines shows that the gain of *PCGNN* methods with respect to *Max power* increases as $\lambda$ increases. This is intuitive since lower $\lambda$ means the deployment has a minor variance in the desired links channel gain and the interference links channel gain, hence power control has less effect. However, the PCGNN methods preserve a significant gain with respect to Max power as $\lambda$ changes. *PCGNN-dD* gain with respect to *Max power* remained relatively constant since the feature selection is based only on the distance metric and independent of $\lambda$. Our proposed *PCGNN-hD* also showed considerable robustness to changes in $\lambda$, considering the tight gap between the red dashed line and the blue continuous line. This means that there is no need to retrain *PCGNN-hD* when

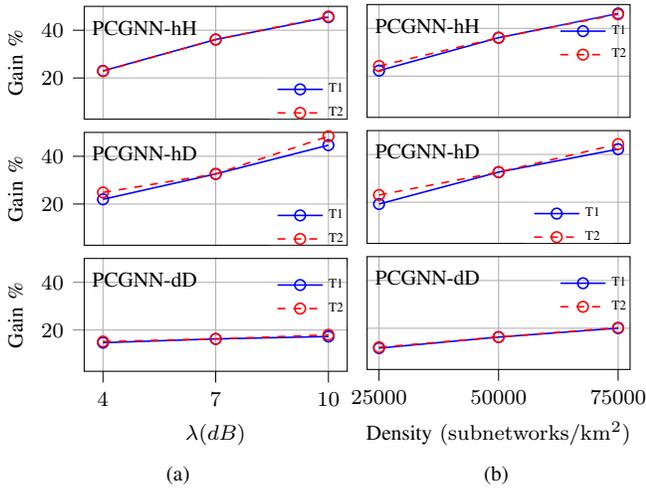

Fig. 5. Robustness to changes in deployment parameters (a) Robustness to variation in shadowing distribution for the 3 *PCGNN* methods *PCGNN-hH*, *PCGNN-hD* and *PCGNN-dD* (b) Robustness to variation in density for the 3 *PCGNN* methods *PCGNN-hH*, *PCGNN-hD* and *PCGNN-dD*. T1- trained and tested with different deployment properties, T2- trained and tested with the same deployment property.

$\lambda$ changes.

2) **Deployment density:** In this case, we changed $N$ in the same factory area thereby reducing or increasing the subnetworks density. In the case of a density of 75000 subnetworks/km$^2$, we increased $N$ to 30. In Figure 5(b), note that; 1) the red dashed lines (legend T2) refer to the gains when the training and testing environments have the same deployment density, and 2) the blue continuous lines refer to the gains when training is performed with the initial environment of deployment density 50000 subnetworks/km$^2$, and testing is done over environments of subnetwork deployment density 25000, 50000, and 75000 subnetworks/km$^2$. We can observe that *PCGNN-hD* remained as robust as *PCGNN-hH*, and the gain with respect to *Max Power* increases as density increases. This is of practical importance as it means that there is no need to retrain the *PCGNN-hD* power control algorithm when there are changes in the number of subnetworks in the factory area.

## V. CONCLUSIONS AND FUTURE WORK

In this paper, we presented a graph neural network approach for optimizing power allocation for industrial wireless subnetworks (IWS) based on a novel input graph model consisting of desired links as node features and interfering link distance as edge features. The approach is motivated by the desire to remove the bottleneck of collecting the interfering link channel gain for the input graph model of the subnetworks during the operational phase. Simulation results show that the proposed PCGNN-hD algorithm performs almost as good as the PCGNN-hH algorithm in terms of achievable average SE, while the latter requires full knowledge of the interfering link channel gain during operations. We also verified the robustness of the proposed solution to variations of shadowing and subnetwork densities.

The solution presented in this paper is intended as a first step towards highly effective low overhead radio resource management for IWS. Future work will further investigate GNN-based solutions for other radio resource management problems, such as dynamic channel allocation, considering additional performance metrics, such as guaranteed minimum transmission rate per subnetwork. In addition, performance will be analyzed with more realistic industrial deployments, including subnetwork mobility (e.g., in mobile robots).